\documentclass[aip,cha,reprint,showkeys,superscriptaddress,nobibnotes, floatfix]{revtex4-1}
\usepackage{soul}
\usepackage{xcolor}
\usepackage{amsmath, amssymb}
\usepackage[pdftex]{graphicx}
\usepackage{epstopdf}

\newcommand{\comment}[1]{}

\begin{document}

\title{Cluster {singularity}: the unfolding of clustering behavior in globally coupled {Stuart-Landau oscillators}}

\author{Felix P. Kemeth}
\author{Sindre W. Haugland}
\affiliation{Physik-Department, Nonequilibrium Chemical Physics, Technische Universit\"{a}t M\"{u}nchen, James-Franck-Str. 1, D-85748 Garching, Germany}
\affiliation{Institute for Advanced Study - Technische Universit\"{a}t M\"{u}nchen, Lichtenbergstr. 2a, D-85748 Garching, Germany}

\author{Katharina Krischer}
\affiliation{Physik-Department, Nonequilibrium Chemical Physics, Technische Universit\"{a}t M\"{u}nchen,
  James-Franck-Str. 1, D-85748 Garching, Germany}
\email[ ]{krischer@ph.tum.de}

\begin{abstract}
  The ubiquitous occurrence of cluster patterns in nature still lacks a comprehensive understanding.
  It is known that the dynamics of many such natural systems is captured by ensembles of Stuart-Landau oscillators.
  Here, we investigate clustering dynamics in a mean-coupled ensemble of such limit-cycle oscillators.
  In particular we show how clustering occurs in minimal networks,
  and elaborate how the observed 2-cluster states crowd when increasing the number of oscillators.
  Using persistence, we discuss how this crowding leads to a continuous transition from balanced cluster states to synchronized solutions via the intermediate unbalanced 2-cluster states.
  These cascade-like transitions emerge from what we call {a} cluster {singularity}.
  At {this codimension-2 point}, the bifurcations of all 2-cluster states collapse and the stable balanced cluster state bifurcates into the synchronized solution supercritically.
  We confirm our results using numerical simulations, and discuss how our conclusions apply to spatially extended systems.
\end{abstract}


\maketitle

\begin{quotation}
Certain swarms of fireflies are known to flash in unison. 
They also sometimes divide into two or more distinct yet internally synchronized groups, flashing with a certain phase lag between the groups.
This is just one example of clustering dynamics in an ensemble of coupled oscillators, as it occurs naturally in many physical systems.
A key problem in the understanding of clustering dynamics is the connection between its occurrence in small and large ensembles.
In other words, is there a universal law governing the arrangement of cluster states, independent of the system size?
This paper partially answers this question and links the phenomenon of clustering in minimal networks of globally coupled limit-cycle oscillators to
clustering in ensembles of infinitely many oscillators. 
We demonstrate that a natural arrangement of such 2-cluster states exists:
When tuning a parameter, a balanced cluster state transitions to synchronized motion via a sequence of intermediate unbalanced cluster states.
Tuning {an} additional {parameter}, this sequence converges to a single point in parameter space where all cluster states {are born} directly {at} the synchronized solution.
We call such a codimension-2 point a cluster singularity.
Singularities of this kind may appear in any symmetrically coupled ensemble of oscillators, and thus
play a crucial role for the understanding of collective behavior in oscillatory systems.
\end{quotation}
Clustering, as discussed above, typically occurs in systems with long-range interactions~\cite{Okuda1993}.
Oscillatory systems with global interactions play a crucial role in the understanding of phenomena observed in nature and technology,
such as the visual perception in the mammalian brain, 
the circadian rhythm in the heart or the behavior of coupled Josephson junctions and electrochemical oscillators.
Even phenomena such as synchronous chirping of crickets, the flashing of fireflies in unison and the synchronous clapping of an audience
can be traced back to the action of a long-range coupling between oscillating units (see Refs.~\cite{Strogatz2000, Pikovsky2015} and references therein).\\
If the coupling between such individual units is weak, each oscillator may be represented by its phase value only,
and the resulting reduced models can be analyzed using powerful approaches as proposed by Okuda~\cite{Okuda1993}, Watanabe and Strogatz~\cite{Watanabe1993,Watanabe1994} or Ott and Antonsen~\cite{Ott2008,Ott2009}.
In many physical systems, however, amplitude effects play a crucial role, and such a reduction is no longer possible.
{Examples include relaxational oscillators with different time scales, such as in neural models~\cite{Ermentrout2010, Golomb1994}. In some settings, clustering can be traced back to the existence of canard explosions~\cite{Yang2000,Rotstein2012}.
In contrast}, close to the onset of oscillations {the oscillatory dynamics} can be reduced
to {a} normal {form description, known} as the Stuart-Landau {oscillator}~\cite{Kuramoto1984,GarciaMorales2012}.
{Due to the universal nature of normal forms, investigations of the dynamics of ensembles of Stuart-Landau oscillators allow general conclusions to be drawn on the dynamics of coupled oscillators independently of the specific model~\cite{GarciaMorales2008, GarciaMorales2008b}.}\\
Systems of globally coupled Stuart-Landau oscillators have been investigated intensively in recent years~\cite{Hakim1992}, revealing phenomena such as collective chaos~\cite{Nakagawa1993}, aging~\cite{Daido2007}, chimera states~\cite{Sethia2014, Kemeth2018}, oscillation death~\cite{Mauparna2014} and clustering~\cite{Aronson1990, Banaji2002, Ku2015, Roehm2018}.\\
{
How cluster states emerge from synchronous motion in ensembles of mean-coupled Stuart-Landau oscillators has been investigated in Ref.~\cite{Banaji2002} using symmetry arguments.
There, the author predicts that the synchronized solution becomes unstable through a transverse bifurcation, consisting of an equivariant pitchfork bifurcation (for an equal number of oscillators),
whose branches represent balanced 2-cluster states,
and a vast amount of transcritical bifurcations, corresponding to unbalanced 2-cluster states with a different number of oscillators in each cluster.
Furthermore, the transition between 2-cluster states has been investigated in Ref.~\cite{Ku2015},
where the authors describe such transitions as hysteretic.\\
In this paper, we investigate the occurrence of 2-cluster states in more detail.
In particular, we resolve the bifurcation scenarios from the synchronized solution to all the different 2-cluster states and demonstrate that the entire bifurcation set can be understood
as the unfolding of a codimension-2 bifurcation, which we call cluster singularity.
Thereby, we start with a system consisting of just two balanced clusters in an ensemble of Stuart-Landau oscillators.
Reducing the equations to the dynamics of just these two clusters, we investigate where such states exist in parameter space for any even number of oscillators.
One has to keep in mind that such a reduction does not reveal any stability properties of the individual clusters transverse to the balanced two-cluster manifold.\\
This can be rectified by increasing the dimension of the system and replacing each of the two original clusters of size $N/2$ with two smaller clusters of size $N/4$ each.
Doing so, we investigate where balanced cluster states are stable in parameter space, and furthermore discuss where 3-1 cluster states exist.\\
Iteratively increasing the dimension of our system allows us to numerically investigate the existence and stability properties of more and more unbalanced 2-cluster states.
In the course of this, we do not only verify the existence of transcritical bifurcations of unbalanced 2-cluster states with the synchronized solution,
but show how such clusters are created in saddle node bifurcations. Unlike in Figure 5 of Ref.~\cite{Banaji2002}, these saddle nodes do not in general contain a stable branch.
The unbalanced cluster states rather get stabilized by additional transverse bifurcations, causing the hysteretic transitions between cluster states.
Furthermore, we describe for the first time the existence of a codimension-2 point which we dub \textit{cluster singularity}.
There, the above-mentioned saddle node and transverse bifurcations collapse onto a single point in phase and parameter space, negating the arrangement of cluster states.
We verify the existence of such a point using numerical simulations, and discuss implications for spatially extended systems.
}
\comment{In particular, we first investigate the stability of 2-cluster solutions in minimal networks of just two and then four oscillators.
We subsequently elaborate how such cluster states bifurcate from the synchronous solution,
and draw connections to previous works on clustering in globally coupled systems.
We then show how new cluster states appear in phase space when increasing the number of oscillators in the network.
Exploiting properties such as persistence yields new insights into how those states are arranged in phase space.
Doing so, we find codimension-2 points which we dub cluster singularities.
There, the balanced cluster states with two clusters of equal size supercritically bifurcate off the synchronized solution.
We continue our considerations with an investigation of how ensembles of Stuart-Landau oscillators behave close to this point,
and what this implies for spatially extended systems.}
{Finally, we} conclude with some open questions concerning clustering in coupled oscillators,
and discuss a few promising directions to address them.\\

\section{Globally coupled Stuart-Landau oscillators}
It has been shown that an oscillatory system close to the onset of oscillations can be reduced to the dynamics of a so-called Stuart-Landau oscillator~\cite{Kuramoto1984,GarciaMorales2012}.
Such analysis has also been extended to cases in which there is a field of oscillators {interacting} with each other.
In particular, for $N$ such systems which interact linearly through the common mean of some of their variables,
one obtains an ensemble of the form
\begin{align}
  \partial_t W_k & = W_k-\left(1+ic_2\right)\left|W_k\right|^2 W_k \nonumber \\
  & + \left(\alpha+i\beta\right)\left(\frac{1}{N}\sum_lW_l-W_k\right),\label{eq:sle}
\end{align}
with $k \in \left\{1,\dots,N\right\}$, the complex amplitude $W_k=W_k\left(t\right) \in \mathbb{C}$, the shear $c_2$, and the coupling constant $\kappa = \alpha+i \beta$, $\alpha,\beta \in \mathbb{R}$~\cite{GarciaMorales2008, GarciaMorales2008b}.
Hereby, the coupling is diffusive in the sense that it vanishes if $W_k=W \; \forall k$~\cite{Aronson1990}.\\
Since we take all oscillators as identical, Eq.~\eqref{eq:sle} is $\Gamma$-equivariant with respect to the symmetric group $\Gamma=\mathbf{S}_N$, that is, with respect to permutations of the indices $k$.
In addition, the dynamics are unchanged under a rotation in the complex plane, $W_k \rightarrow W_ke^{i\phi}$, and under the reflection $W_k \rightarrow \overline{W}_k, c_2\rightarrow -c_2, \beta\rightarrow -\beta$, with the bar indicating complex conjugation.\\
Eq.~\eqref{eq:sle} can be seen as a normal form for oscillatory systems with a quickly diffusing variable or a coupling through e.g. a gas phase~{\cite{Falcke1993, Kim2001, Bertram2003}}.
There is a vast number of different dynamical states that can be observed in such an ensemble,
including synchronized motion and splay states~\cite{Hakim1992}, cluster states~\cite{Okuda1993, Banaji2002, Ku2015},
chaotic dynamics~\cite{Nakagawa1993,Nakagawa1994,Nakagawa1995} and chimera states~\cite{Sethia2014}.\\
For the synchronized solution, in which all $N$ oscillators move in phase, and the splay-state, in which all oscillators are frequency-synchronized but with a phase difference of $2\pi/N$,
the respective stability boundaries can be obtained analytically, see for example Ref.~\cite{Ku2015}.
For chaotic or chimera-like dynamics, however, only numerical approaches exist.\\
In this article, we restrict our analysis to solutions in which the ensemble splits into just two groups with each group being internally synchronized, also called a 2-cluster state~\cite{Banaji2002}.
Let $\epsilon=N_1/N$ denote the fraction of oscillators in cluster $1$ and $\left(1-\epsilon\right)=N_2/N$ the fraction of oscillators in cluster $2$, then Eq.~\eqref{eq:sle} reduces to
\begin{align*}
  \partial_t W_1 &= W_1 - \left(1+ic_2\right)\left|W_1\right|^2 W_1 \\
  & + \left(\alpha+i\beta\right)\left(1-\epsilon\right)\left( W_2 - W_1\right)\\
  \partial_t W_2 &= W_2 - \left(1+ic_2\right)\left|W_2\right|^2 W_2\\
  & + \left(\alpha+i\beta\right)\epsilon\left(W_1 -  W_2\right),
\end{align*}
with $\left(W_1,W_2\right)\in \mathbb{C}^2\cong\mathbb{R}^4$.
{
Here, we can in fact regard the two clusters as just two oscillators with weights $\epsilon$ and $1-\epsilon$, respectively. The governing equations in each case are the same.
One has to keep in mind, however, that in such a reduced system the individual clusters cannot break up, and thus no information about transversal stability can be obtained.}\\
Following Ref.~\cite{Pikovsky2001_a}, the dynamics can be further reduced by exploiting the rotational invariance $W_k \rightarrow W_ke^{i\phi}$ mentioned above and introducing polar coordinates
${W_1=R_1e^{i\phi_1}}$, ${W_2=R_2e^{i\phi_2}}$ and the phase difference ${\Delta \phi = \phi_2-\phi_1}$.
This yields the reduced equations
\begin{align}
  \partial_t R_1 & = R_1-R_1^3 + \alpha\left(1-\epsilon\right)\left(R_2\cos\left(\Delta \phi\right)-R_1\right) \nonumber \\
  & + \beta\left(1-\epsilon\right) R_2\sin \left(\Delta \phi\right)\label{eqn:tsl_r1}\\
  \partial_t R_2 &= R_2-R_2^3  + \alpha\epsilon\left(R_1\cos\left(\Delta \phi\right)-R_2\right) \nonumber \\
  & - \beta\epsilon R_1\sin \left(\Delta \phi\right)\label{eqn:tsl_r2}\\
  \partial_t\Delta\phi & = -c_2\left(R_1^2-R_2^2\right)+\beta\left(1-2\epsilon\right) \nonumber\\
                 & + \beta\cos\left(\Delta\phi \right)\left(\frac{\left(1-\epsilon\right)R_2}{R_1} -\frac{\epsilon R_1}{R_2}\right) \nonumber \\
  & - \alpha\sin\left(\Delta\phi\right)\left(\frac{\left(1-\epsilon\right)R_2}{R_1}+\frac{\epsilon R_1}{R_2}\right)\label{eqn:tsl_dp0},
\end{align}
describing the dynamics in a three-dimensional phase space $\mathbb{R}^+\times \mathbb{R}^+ \times \mathbf{T}$ with $\mathbf{T} = \mathbb{R}/2\pi \mathbb{Z}$.
Note that the roots of the reduced model, Eqs.~\eqref{eqn:tsl_r1} to~\eqref{eqn:tsl_dp0}, correspond to sinusoidal oscillations in the full system {Eq.~\eqref{eq:sle}}.
\section{Balanced 2-Cluster States}
First, we restrict our analysis to 2-cluster states with an equal number of oscillators in each cluster, $N_1=N_2=N/2$, that is for $\epsilon=1/2$.
Then Eqs.~\eqref{eqn:tsl_r1} to~\eqref{eqn:tsl_dp0} in fact describe the dynamics of just two identical oscillators, {each oscillator representing one cluster}.
If the coupling is weak, that is if {$\lVert\alpha+i\beta\rVert<<1$}, then the amplitudes relax to $R_1=1$, $R_2=1$ on a fast time scale, and the dynamics can be described,
to a good approximation, by the phase equation only~\cite{Kuramoto1984}.
Therefore, with the adiabatic approximation $R_1=R_2=1$ the system, Eqs.~\eqref{eqn:tsl_r1}~to~\eqref{eqn:tsl_dp0}, reduces to
\begin{equation}
  \partial_t\Delta\phi = - \alpha\sin\left(\Delta\phi\right),
\end{equation}
and thus to a sine-coupled system~\cite{Okuda1993}.
Notice that it has two fixed points at $\Delta\phi=0$ and $\Delta\phi=\pm \pi$, and therefore only the in-phase and anti-phase solutions exist for weak coupling.
Important findings for larger ensembles of limit-cycle oscillators with weak coupling, that is for globally coupled phase oscillator systems,
are summarized in~\cite{Kuramoto1987} and~\cite{Pikovsky2015}.\\
The fixed points of the 2-oscillator system, Eqs.~\eqref{eqn:tsl_r1}~to~\eqref{eqn:tsl_dp0} with $\epsilon=1/2$, can be determined as
\begin{itemize}
\item $\vec{u}_s = \left(R_1,R_2,\Delta \phi\right)^\mathrm{T} = \left(1,1,0\right)$,
  the synchronized solution in which both oscillators have amplitude equal to one and phase difference zero,
\item $\vec{u}_a = \left(\sqrt{1-\alpha},\sqrt{1-\alpha},\pi\right)$,
  the anti-phase solution corresponding to a splay state for two oscillators,
\item and two symmetry-broken solutions $\vec{u}_{1,2}$ (see Appendix A for their derivation).
\end{itemize}
Note that the anti-phase solution $\vec{u}_a$ exists only for parameter values $\alpha<1$,
and that the locations in phase space of both the synchronized and anti-phase solutions are independent of the parameters $\beta$ and $c_2$.
Their stability boundaries can be determined by calculating the eigenvalues of the Jacobian evaluated at these solutions,
and are given by the Benjamin-Feir condition~\cite{Nakagawa1993, Hakim1992,Benjamin1967},
\begin{equation}
  \alpha^2+\beta^2 + 2\alpha+2\beta c_2 =0,
  \label{eq:bf_synch}
\end{equation}
for the synchronized solution and by
\begin{equation}
  \alpha^2 + \beta^2 - 2\left(1-\alpha\right)\left(\alpha+\beta c_2\right) =0
  \label{eq:bf_asynch}
\end{equation}
for the anti-phase solution.
It is worth mentioning that these stability boundaries are independent of the number of oscillators in the ensemble~\cite{Nakagawa1993}.
The stability diagram of these two solutions for $c_2=0$ is depicted in Fig.~\ref{fig:0}(a).
There, one can observe that the synchronized solution is stable for positive $\alpha$  values,
and either loses one stable direction at the solid blue curve and subsequently a second stable direction at the dotted blue curve,
or two stable directions through a Hopf bifurcation at the horizontal blue lines.
Analogously, the anti-phase solution is unstable for $\alpha>0.5$, and gains one stable direction either at the dotted red curve and subsequently becomes stable at the solid red curve,
or it becomes directly stable through a Hopf bifurcation at the horizontal red line where it gains two stable eigendirections.
Note that there exist two regions in which the two solutions are bistable.\\
The asymmetric solutions $\vec{u}_{1,2}$ (see Appendix A for their derivation) have the property that the amplitudes of the two oscillators differ
and the oscillators have a {phase difference between $0$ and $\pi$}, see also Ref.~\cite{Roehm2018}.
\begin{figure}[ht]
  \centering
  \includegraphics{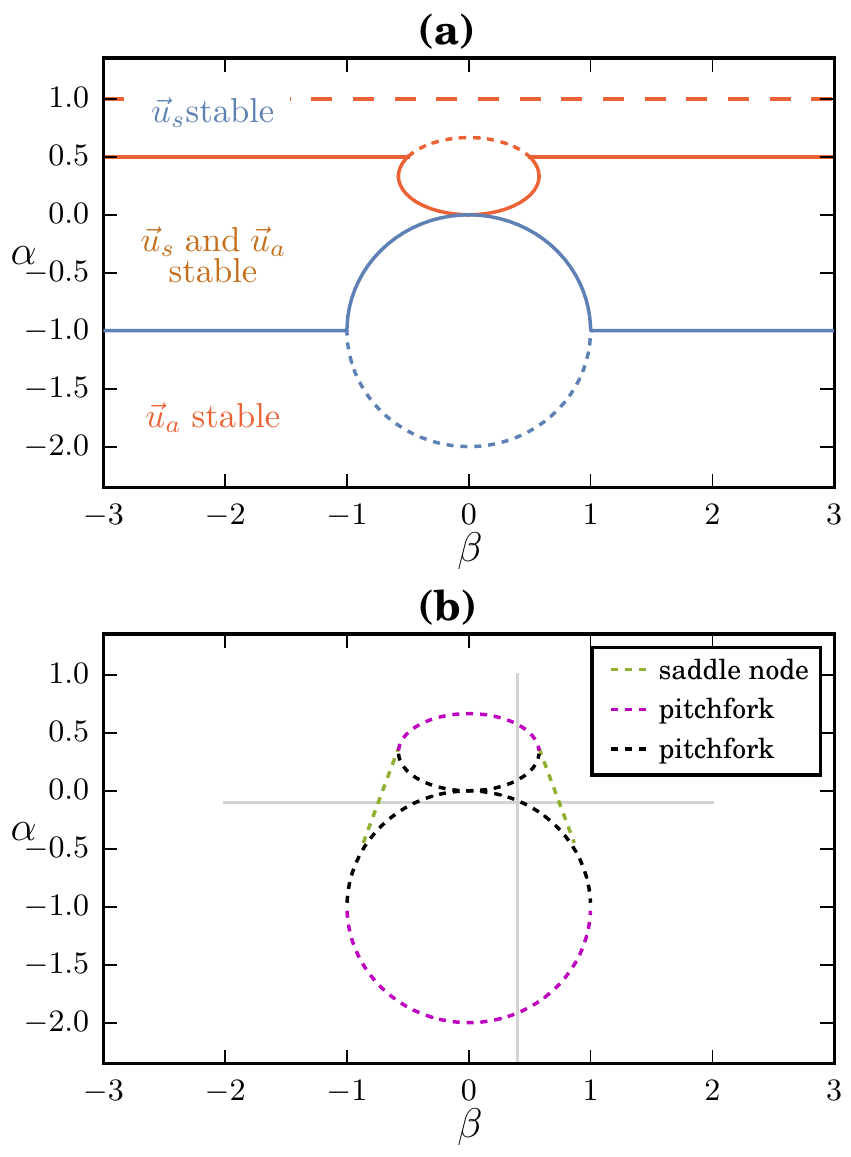}
  \caption{(a) Stability {boundaries} of the synchronized solution $\vec{u}_s$ (blue) and anti-phase solution $\vec{u}_a$ (red) for $c_2=0$.
      From negative to positive $\alpha$, the synchronized solution $\vec{u}_s$, having two unstable directions,
      gains a stable direction at the dotted blue curve, and eventually becomes stable at the solid blue curve {(Eq.~\eqref{eq:bf_synch})}.
      At the horizontal blue line, the real parts of two complex conjugate eigenvalues of the Jacobian cross the imaginary axis, indicating a Hopf bifurcation.
      In contrast, the anti-phase solution $\vec{u}_a$ loses two stable directions at the horizontal solid red line,
      whereas it looses one unstable direction at the solid red curve {(Eq.~\eqref{eq:bf_asynch})} and another at the dotted red curve.
      This fixed-point solution eventually disappears at $\alpha=1$, indicated by the dashed red line, where the amplitudes $R_1$ and $R_2$ vanish.
      (b) {Boundaries of the parameter windows in which the asymmetric solutions $\vec{u}_{1,2}$ exist for $c_2=0$.}
      Hereby, solution $\vec{u}_1$ exists for parameter values between the saddle-node bifurcations, indicated by the dotted green lines, and between the dotted magenta curves symbolizing pitchfork bifurcations.
      The boundaries of the solutions $\vec{u}_2$ are again the dotted green lines, and the pitchfork {bifurcations} indicated by the dotted black lines.
      The gray lines indicate the one-parameter continuation cuts shown in Fig.~\ref{fig:1_1ab}.
  }
  \label{fig:0}
\end{figure}
\comment{Note that, since our original system is $\mathbf{S}_N$-equivariant, every symmetry-broken state must belong to a subgroup of $\mathbf{S}_N$.
In particular, for any 2-cluster states $\vec{u}_{\text{cl}}$, the isotropy subgroup $\Sigma_{\vec{u}_{\text{cl}}}$ has the form $\mathbf{S}_{N_1}\times\mathbf{S}_{N_2}\subseteq \mathbf{S}_N$.}
{Considering the symmerty of the cluster solutions, we note that for any 2-cluster state $\vec{u}_{\text{cl}}$, the isotropy subgroup $\Sigma_{\vec{u}_{\text{cl}}}$ has the form $\mathbf{S}_{N_1}\times\mathbf{S}_{N_2}\subseteq \mathbf{S}_N$.}
Furthermore, let $\left\{\gamma \in \mathbf{S}_N\backslash \Sigma_{\vec{u}_{\text{cl}}}\right\}$ denote the set of operations {in $\mathbf{S}_N$} not included in the isotropy subgroup of $\vec{u}_{\text{cl}}${. Then} there exist $N_{\vec{u}_{\text{cl}}}=\left|\left\{\gamma \in \mathbf{S}_N\backslash \Sigma_{\vec{u}_{\text{cl}}}\right\}\right|$ cluster solutions related to ${\vec{u}_{\text{cl}}}$, which form
the so-called group orbit of ${\vec{u}_{\text{cl}}}$~\cite{Golubitsky2003}. For example, considering the 2-cluster solution ${\vec{u}_{\text{cl}}}=\left(R_1,R_2,\Delta \phi\right)$,
then $\vec{\tilde{u}}_{\text{cl}}=\left(R_2,R_1,-\Delta \phi\right)$ is also a 2-cluster solution and $\vec{u}_{\text{cl}}$, $\vec{\tilde{u}}_{\text{cl}}$ belong to the same group orbit.\\
\begin{figure}[ht]
  \centering
  \includegraphics{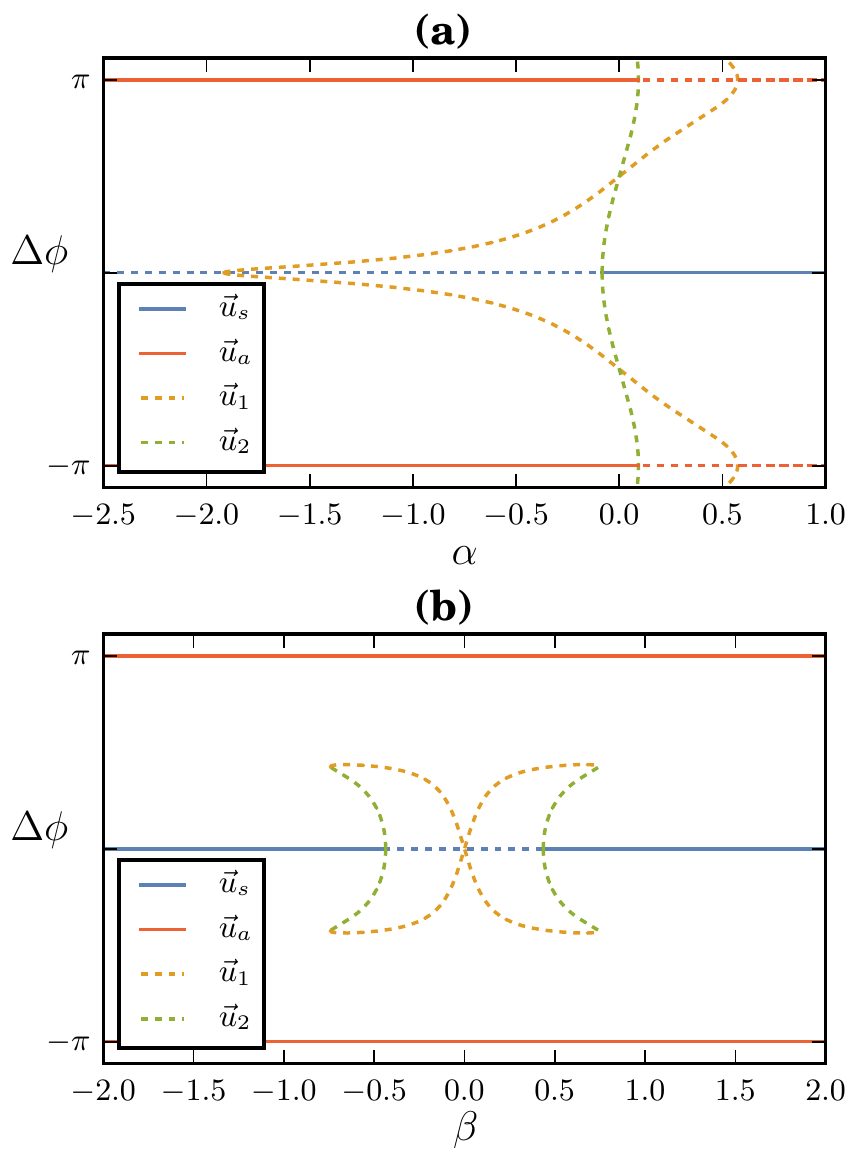}
  \caption{Continuation of the fixed-point solutions in the reduced model of four mean-coupled oscillators with AUTO, (a) along the vertical cut shown in Fig.~\ref{fig:0}(b) with $\beta=0.4$,
    (b) along the horizontal cut in Fig.~\ref{fig:0}(b) with $\alpha=-0.1$. Stable solutions are shown as solid curves, whereas unstable branches are indicated by dashed curves.
  Note that the branches at $\pm \pi$ symbolize the same solution $\vec{u}_a$.}
  \label{fig:1_1ab}
\end{figure}
In the case of no shear, $c_2=0$, asymmetric solutions $\vec{u}_{1,2}$ bifurcate off the anti-phase solution for large $\alpha$ values through pitchfork bifurcations {(see the dotted magenta and black curves for $\alpha>0$ in Fig.~\ref{fig:0}(b))}.
For smaller $\alpha$ values, they either bifurcate in a saddle-node with each other (see the dotted green line in Fig.~\ref{fig:0}(b))
or with the synchronized solution through pitchfork bifurcations (see the dotted magenta and black curves for $\alpha<0$ in Fig.~\ref{fig:0}(b)).
That means, $\vec{u}_1$ (to be more precise, the two elements of the group orbit of $\vec{u}_1$) exists between the dotted magenta curves (pitchforks) and the dotted green lines (saddle-nodes),
and the solution  $\vec{u}_2$ exists in the two small regions between the dotted black curves (pitchforks) and the dotted green lines (saddle-nodes).\\
The continuation of the different solutions along one-parameter cuts (as indicated by the gray lines in Fig.~\ref{fig:0}(b)) is shown in Fig.~\ref{fig:1_1ab},
where each solution is represented by its $\Delta \phi$ variable.
This means that the synchronized solution is located at $\Delta \phi=0$ and the anti-phase solution at $\pm\pi$.
In Fig.~\ref{fig:1_1ab}(a), $\beta=0.4$ is held fixed and the synchronous solution is continued with increasing $\alpha$.
There, one observes a subcritical pitchfork at which the solution branches of $\vec{u}_1$ are created and $\vec{u}_s$ gains an additional stable direction.
After another pitchfork the synchronized solution becomes stable and the two branches of $\vec{u}_2$ are born.
Those two branches then reach the anti-phase solution $\vec{u}_a$ where they get destroyed in a pitchfork bifurcation, rendering $\vec{u}_a$ unstable.
Finally, the branches of $\vec{u}_1$ bifurcate with $\vec{u}_a$ in another pitchfork, adding the second unstable direction.\\
Fixing $\alpha$ at $\alpha=-0.1$ and starting from negative $\beta$ values,
we observe two saddle-node bifurcations, each involving one branch of $\vec{u}_1$ and $\vec{u}_2$, respectively.
The branches of $\vec{u}_2$ are annihilated through a subcritical pitchfork at the synchronized solution.
Due to the symmetry $\beta\rightarrow -\beta$, this scenario holds also for positive $\beta$ values.
Note that there is no bifurcation at $\beta=0$.
The crossing of the two branches of $\vec{u}_1$ is no real crossing but is a consequence of the projection of the solutions onto the $\Delta \phi$ variable.\\
The stability of the asymmetric solutions $\vec{u}_{1,2}$ can further be investigated using the eigenvalues of the Jacobian.
Doing so, we find that $\vec{u}_2$ is always unstable in the regarded parameter regimes, but $\vec{u}_1$ is stable for certain ranges of $\alpha$, $\beta$ and $c_2$.
In particular, the number of unstable eigendirections of $\vec{u}_1$ in the $\alpha$-$\beta$ parameter plane is shown in Fig.~\ref{fig:1} for different values of $c_2$.
\begin{figure*}[ht]
  \centering
  \includegraphics{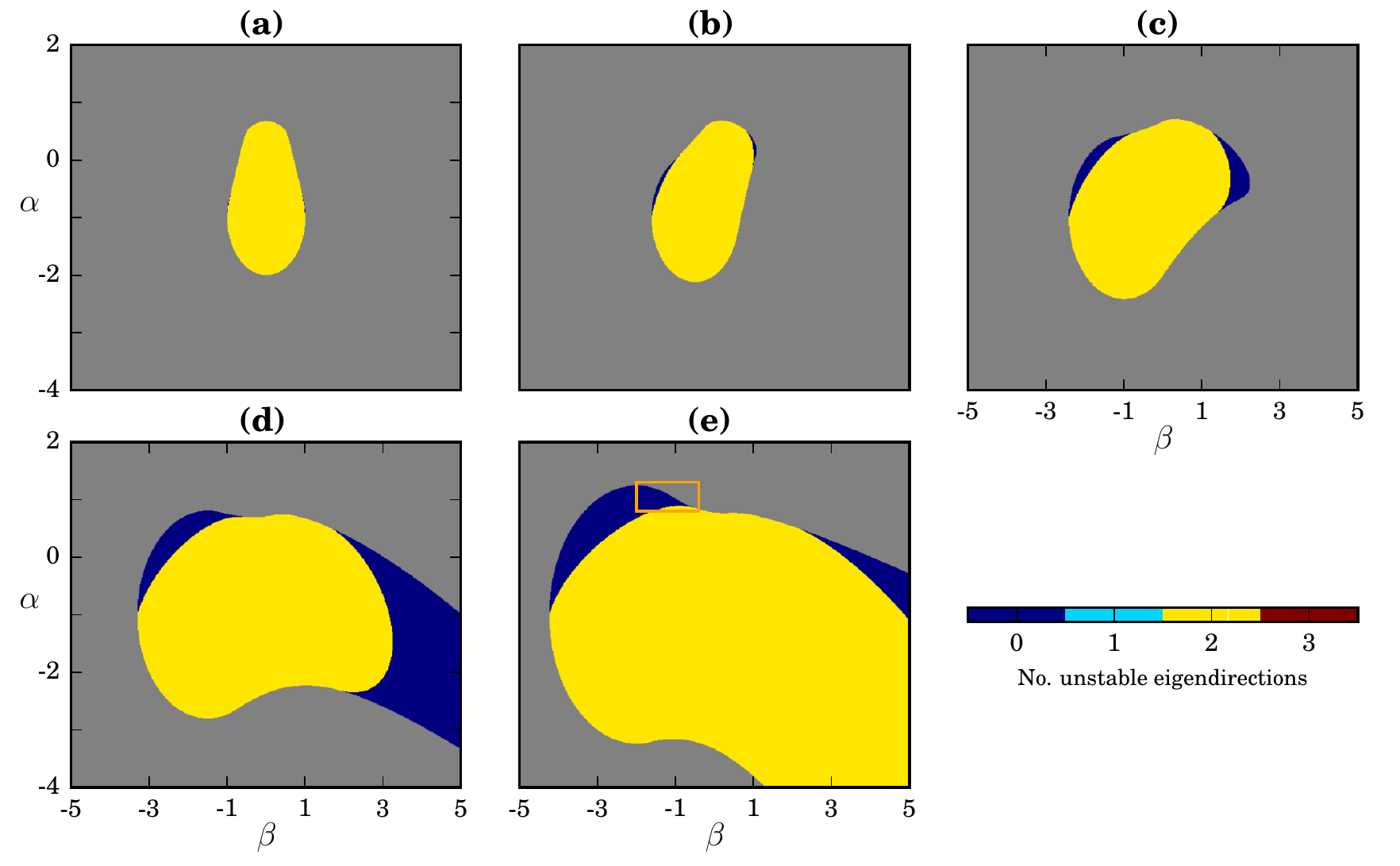}
  \caption{
    {Parameter windows in the $\beta$-$\alpha$ plane for which the asymmetric solution $\vec{u}_1$ exists in an ensemble of two coupled Stuart-Landau oscillators for (a) $c_2=0$, (b) $c_2=0.5$, (c) $c_2=1$, (d) $c_2=1.5$ and (e) $c_2=2.0$.}
    \comment{Stability of the asymmetric solution $\vec{u}_1$ in an ensemble of two coupled Stuart-Landau oscillators for (a) $c_2=0$, (b) $c_2=0.5$, (c) $c_2=1$, (d) $c_2=1.5$ and (e) $c_2=2.0$.}
    The color encodes the number of unstable eigendirections whereas gray indicates the parameter range where the asymmetric solution does not exist.
    Since the system is invariant under $c_2\rightarrow -c_2$, $\beta \rightarrow -\beta$ plots of only positive $c_2$ values are shown here.
    A magnification of the orange window is shown in Fig.~\ref{fig:2}(a).}
  \label{fig:1}
\end{figure*}
Note that even for $c_2=0$ small regions exist in which $\vec{u}_1$ is stable, but these regions grow for larger absolute values of $c_2$, see Figs.~\ref{fig:1}(b)-(e).
The stable 2-cluster states lose stability through a Hopf bifurcation, leading to the yellow patches in Fig.~\ref{fig:1} with two unstable directions.
The Hopf bifurcation is supercritical for small $c_2$, creating stable periodic orbits~\cite{Nakagawa1993}.
Due to the symmetries in the equations, as discussed above, the bifurcation diagrams are symmetric under a simultaneous exchange $\beta\rightarrow -\beta$, $c_2\rightarrow -c_2$,
leading to the left-right symmetry in Fig.~\ref{fig:1}(a).
{\section{Stability of 2-Cluster States}}
The reduced model, Eqs.~\eqref{eqn:tsl_r1} to~\eqref{eqn:tsl_dp0},
and the eigenvalues of its Jacobian offer only limited information about the stability of 2-cluster states.
This becomes obvious when considering the different ways in which 2-cluster solutions can bifurcate:
either on or transverse to the cluster manifold.
That is, either the two cluster clumps each remain together, but their relative locations change, or one {or both} of the two clusters split up into an arbitrary number of subclusters.
The stability properties of the former are fully captured by the reduced equations, but since the latter cannot happen in the reduced model,
we can draw no conclusions about stability transverse to the cluster manifold.\\
In order to rephrase these arguments, we follow Ref.~\cite{Ku2015} and describe the stability of 2-cluster states by two kinds of Lyapunov exponents,
the \textit{cluster integrity exponents} $\lambda_{CI}^{\sigma}$,
and the \textit{cluster system orbit stability exponents} $\lambda_{SO}$.
The latter describe the stability along the 2-cluster manifold and constitute 3 real numbers.
The former, the cluster integrity exponents $\lambda_{CI}^{\sigma}$, describe the internal stability of each cluster $\sigma$.
Due to the symmetries of the oscillator ensemble, each $\lambda_{CI}^{\sigma}$ is degenerate in the sense that they consist of $2N_\sigma-2$ equal real values,
with $N_\sigma$ being the number of oscillators in cluster $\sigma$.\\
We extend our considerations by regarding four instead of two equally sized clusters.
By again introducing polar coordinates, we obtain the dynamics of the four amplitudes $R_k$, $k=1,\dots,4$,
and three phase differences $\Delta \phi_{k1} = \phi_{k}-\phi_1$, $k=2,3,4$.
One thus has dynamics in a seven-dimensional phase space $\mathbb{R}_+^4\times \mathbf{T}^3$,
which is equivalent to the dynamics of four coupled Stuart-Landau oscillators, each representing one cluster~\cite{Kemeth2018}.
Note that the asymmetric solutions $\vec{u}_{1,2}$ now correspond to cluster solutions with two oscillators in each cluster.
Due to the new dimensions in phase space, however, the stability of those 2-2 cluster solutions might differ from the stability of the $\vec{u}_{1,2}$ solutions in the 2-oscillator ensemble.
This can be visualized by evaluating the Jacobian of the 4-oscillator system at the 2-2 cluster solutions and investigating the number of
eigenvalues with positive real parts.
\begin{figure}[ht]
  \centering
  \includegraphics{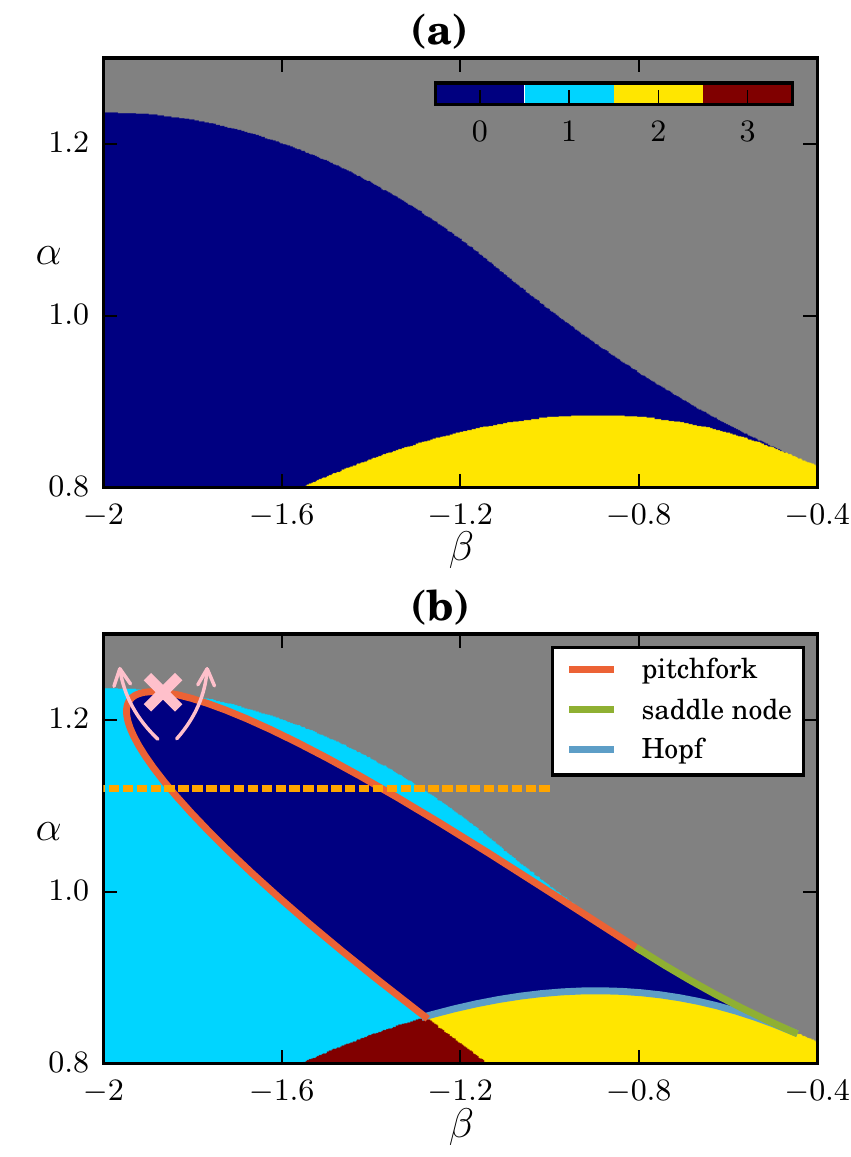}
  \caption{{Parameter window in the $\beta$-$\alpha$ plane for which the balanced cluster solution exists in an ensemble of (a) two and (b) four mean-coupled Stuart-Landau oscillators for $c_2=2$.}
    \comment{Stability of the balanced cluster solution in an ensemble of (a) two and (b) four mean-coupled Stuart-Landau oscillators for $c_2=2$.}
    The parameter range corresponds to the orange window indicated in Fig.~\ref{fig:1}(e).
    As in Fig.~\ref{fig:1}, the color encodes the number of unstable eigendirections, and gray indicates that the balanced cluster solution does not exist for this set of parameters.
    {In particular, in the gray region the synchronized solution is stable.}
    The bifurcation curves plotted in {Fig.~\ref{fig:2}(b)} were obtained using AUTO and symbolize a supercritical Hopf bifurcation curve (blue),
    a pitchfork bifurcation curve (red) and a saddle-node bifurcation curve (green).
    The position of the cluster singularity (as discussed in the subsequent sections of this article) is indicated by the pink x.}
  \label{fig:2}
\end{figure}
For the parameter window shown {for the 2-oscillator system} in Fig.~\ref{fig:2}(a) (corresponding to the orange window in Fig.~\ref{fig:1}(e)),
the stability of the 2-2 cluster solution in the 4-oscillator system for $c_2=2$ is shown in Fig.~\ref{fig:2}(b).
There, one can observe that the parameter range in which the 2-2 cluster solution is stable is smaller than the parameter regime in which the $\vec{u}_{1}$ solution is stable
in the 2-oscillator ensemble.
Using the numerical continuation software AUTO, one can further observe that the 2-2 cluster solution may either become unstable through a pitchfork (red curve in Fig.~\ref{fig:2}(b)),
a saddle-node (green curve in Fig.~\ref{fig:2}(b)), or through a supercritical Hopf bifurcation (blue curve in Fig.~\ref{fig:2}(b)).\\
Note that for the 2-2 cluster in the 4-oscillator ensemble, we have $2\cdot 2-2=2$ cluster integrity exponents of one cluster,
$2$ cluster integrity exponents of the other cluster and 3 cluster system orbit stability exponents, and thus 7 exponents in total.
However, the degeneracy of the cluster integrity exponents implies that if a 2-2 cluster state is stable for certain parameters in the 4-oscillator ensemble,
then any $N/2$-$N/2$ cluster state with an even-valued $N\geq 4$ is stable for such parameters.
Therefore, the {dark blue region} shown in Fig.~\ref{fig:2}(b) indicates where balanced 2-cluster states (that is, cluster states with the same number of oscillators in each of the two clusters) for any even number of oscillators (larger than four) are stable.\\
\begin{figure}[ht]
  \centering
  \includegraphics{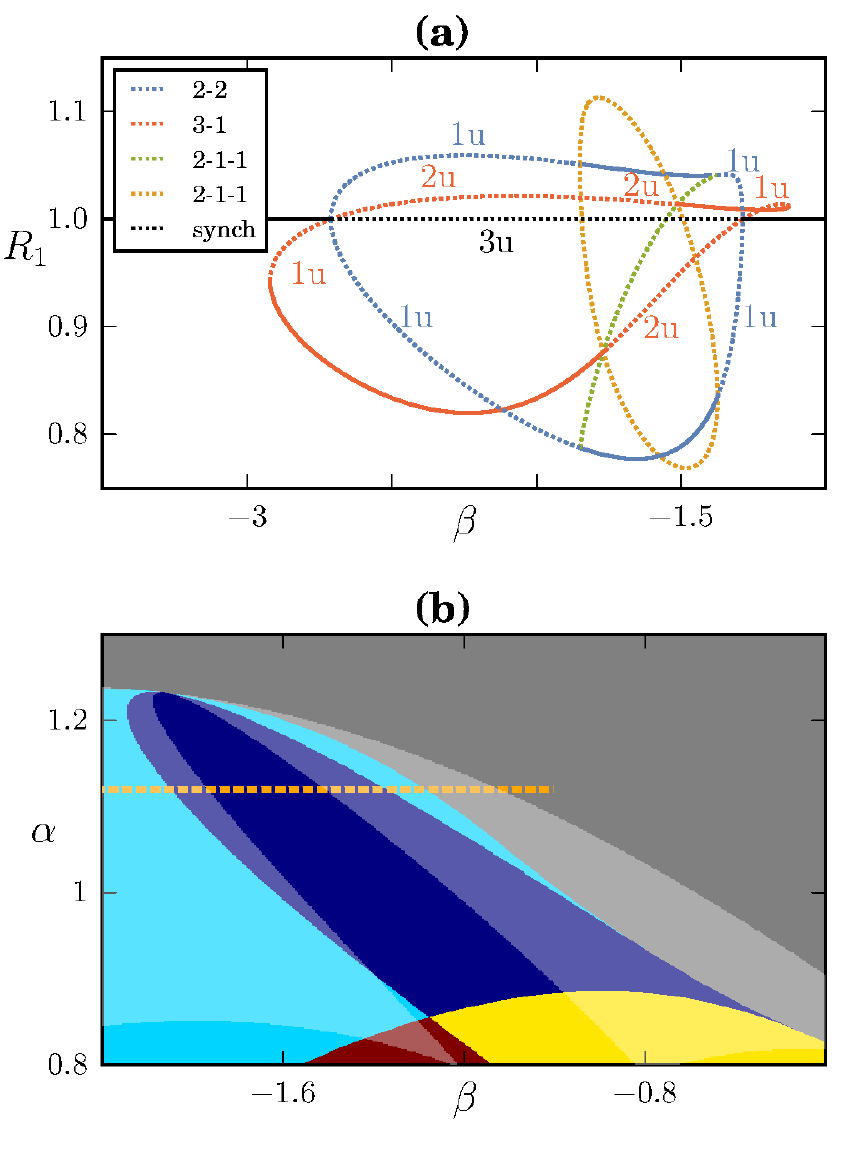}
  \caption{(a) Continuation of the cluster solutions in {an} ensemble of four coupled
    Stuart-Landau oscillators for $\alpha = 1.12$ and $c_2 = 2$, as indicated by the dashed
    orange line in Fig.~\ref{fig:2}(a) but extended to more negative $\beta$ values, using AUTO.
    Dotted curves indicate {unstable solutions}, whereas
    solid lines represent attracting solutions. The numbers indicate the number of unstable eigendirections.
    (b) {Parameter window in the $\beta$-$\alpha$ plane for which the balanced cluster solution exists in an ensemble of four mean-coupled Stuart-Landau oscillators for $c_2=2$.}
    \comment{Stability of the balanced cluster solution in an ensemble of four mean-coupled Stuart-Landau oscillators for $c_2=2$.}
    The parameter range corresponds to the orange window indicated in Fig.~\ref{fig:1}(e).
    As in Fig.~\ref{fig:1}, the color encodes the number {of} unstable eigendirections, and gray indicates that the balanced cluster solution does not exist for this set of parameters.
    {In addition to the stability of the balanced cluster state, as also shown in Fig.~\ref{fig:2}(b), the parameter regions in which the 3-1 cluster states are stable are indicated by means of white-shaded patches.}
    \comment{The regions in which the 3-1 cluster states are stable are indicated by the white-shaded patches.}
  {Note that the $\beta$-range in (b) differs from the one in (a).}}
  \label{fig:3}
\end{figure}
The question remains how this 2-2 cluster state bifurcates from or to the synchronized solution.
In order to investigate this, we look at a cut in parameter space (indicated by the dashed orange line in Fig.~\ref{fig:2}(b)) and continue the 2-2 cluster solution by varying $\beta$ for fixed $\alpha$ and $c_2$.
The amplitude of one cluster, $R_1$, from an exemplary continuation with AUTO is depicted in Fig.~\ref{fig:3}(a).
There, the 2-2 cluster state is indicated by blue curves, which are dotted when this cluster solution is unstable and solid when the 2-2 cluster is stable.
Note that there exist two varieties of the 2-2 cluster state, as indicated by the upper and lower blue line in Fig.~\ref{fig:3}(a).
These two solutions differ in their respective value of the amplitude $R_1$, but are both stable for the same $\beta$ values.
Moreover, they belong to the same group orbit (as they can be transformed into one another by interchanging the oscillators).
Furthermore{,} one can observe that the 2-2 cluster states bifurcate into 2-1-1 cluster states (green and orange curves in Fig.~\ref{fig:3}(a)),
which are unstable for the parameter values regarded here.
These 2-1-1 branches, however, subsequently bifurcate into 3-1 cluster branches (red curve in Fig.~\ref{fig:3}(a)), which shows two stable regions.
In contrast to the stable 2-2 cluster solutions, the stable 3-1 cluster solutions do not belong to the same group orbit,
but correspond to a state in which the
cluster with {three} oscillators has an amplitude $R_1>1$ (for $-1.6 \leq \beta \leq -1.2$) and to a state in which the cluster with three oscillators has an amplitude $R_1<1$ (for $-2.9\leq \beta \leq -1.7$).
Note that at high {respectively} low $\beta$ values the stable 3-1 cluster states {get destroyed} through saddle-node bifurcations.
In addition, the {3-1} cluster states bifurcate with the synchronized solution
(black line in Fig.~\ref{fig:3}(a)) through a transcritical bifurcation,
a scenario which has already been described in Refs.~\cite{Ashwin1992,Banaji2002}.
This is in contrast to the ensemble of symmetrically related 2-2 cluster states, which bifurcate off the synchronized solution through equivariant pitchfork bifurcations at the same points.
The overlapping {regions of stability of the} 2-2 and 3-1 cluster solutions shown in Fig.~\ref{fig:3}(a) also explain the hysteretic transitions from balanced cluster states to homogeneous oscillations and vice versa{, as explained in more detail in the following paragraphs}.
The regime in which the 3-1 cluster solution is stable is shown as a shaded region in Fig.~\ref{fig:3}(b).\\
\begin{figure*}[ht]
  \centering
  \includegraphics{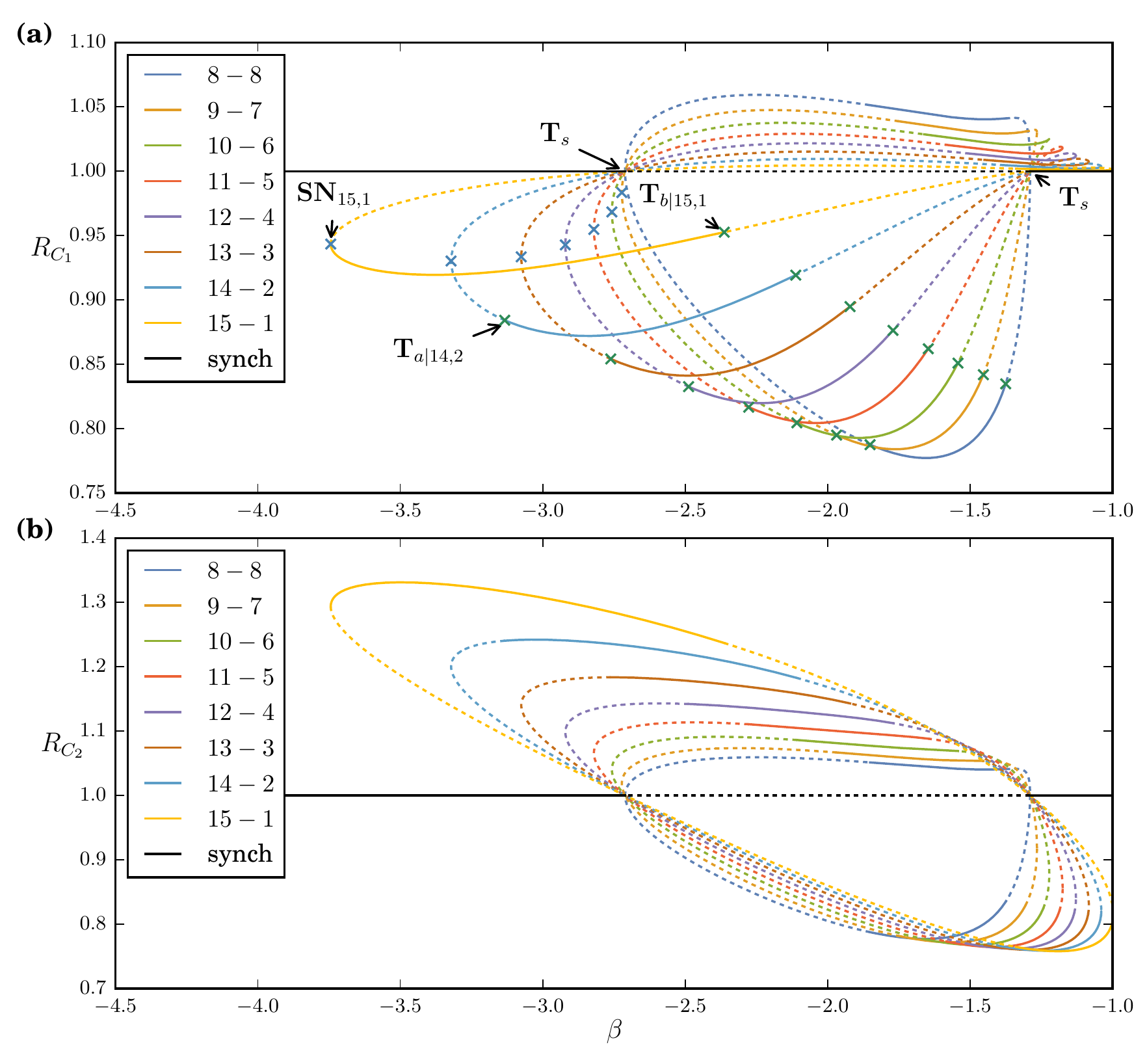}
  \caption{Cluster solutions in an ensemble of 16 coupled Stuart-Landau oscillators obtained using AUTO.
    Shown is (a) the amplitude of the larger cluster, $R_{C_1}$, as a function of the bifurcation parameter $\beta$, as indicated by the dashed orange line in Fig.~\ref{fig:3}(b),
    and (b) the amplitude of the smaller cluster, $R_{C_2}$. {The blue crosses indicate the positions of the saddle nodes $\mathbf{SN}_{N-x, x}$ creating the unbalanced cluster solutions.
      The green crosses indicate the positions of the transverse bifurcations stabilizing the unbalanced cluster solutions, $\mathbf{T}_{a | N-x, x}$, and destabilizing the unbalanced cluster solutions, $\mathbf{T}_{b | N-x, x}$. Due to space limitations, only 2-cluster branches are shown, and the bifurcation points are indicated only for small $\beta$ values.}
    Other parameter values are $c_2=2$ and $\alpha=1.12$.}
  \label{fig:4}
\end{figure*}
\begin{figure*}[ht]
  \centering
  \includegraphics{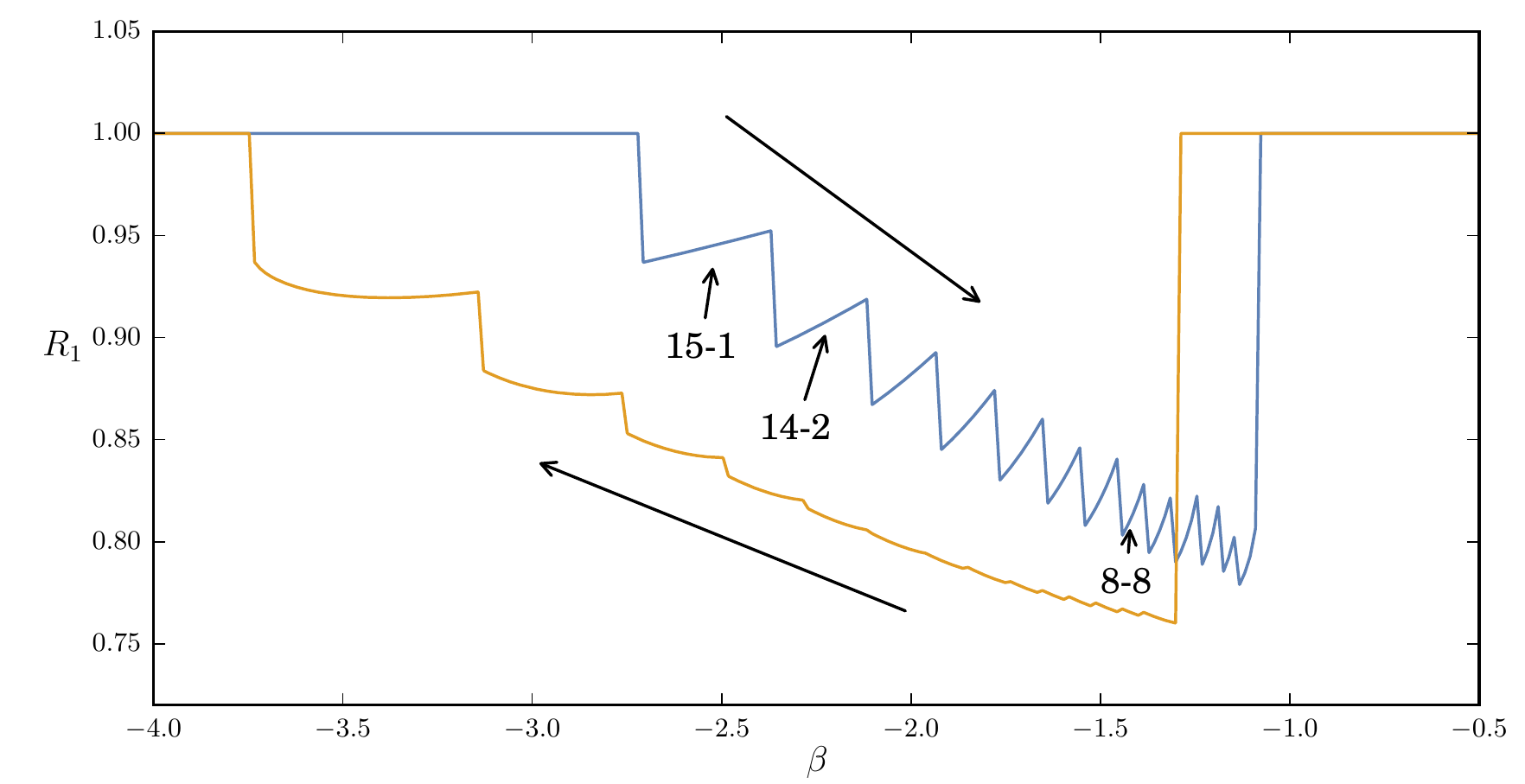}
  \caption{The amplitude of one of 16 oscillators as a function of the parameter $\beta$ when starting from the stable synchronized solution, increasing $\beta$ (blue curve)
    and subsequently reducing $\beta$ (orange curve).
    Due to the addition of finite noise in the numerical simulations when increasing $\beta$, not all of the densely located cluster states close to $\beta_2$ are resolved.}
 \label{fig:hyst_trans}
\end{figure*}
The analysis from above can easily be extended to larger ensembles of oscillators.
In particular, we now consider $16$ oscillators and investigate the clustering behavior along the parameter cut indicated by the dashed orange line in Fig.~\ref{fig:2}(b).
That is, we fix $c_2=2$ and $\alpha=1.12$ and vary $\beta$.
We do this by simulating the full model starting from the balanced 8-8 cluster solution, and increasing $\beta$ stepwise.
Over the course of this increase, the system goes from a stable 8-8 solution to a stable 9-7 solution, 10-6 solution and so forth until it settles on the synchronized solution.
In this way we obtain solutions for every cluster distribution, which we then use for continuation with AUTO.
The continuation curves we obtain this way are depicted in Fig.~\ref{fig:4}(a), where again the amplitude of the cluster with the larger number of oscillators, $R_{C_1}$,
is shown as a function of the continuation parameter $\beta$.
The amplitudes of the smaller cluster for the same parameter window are shown in Fig.~\ref{fig:4}(b).
Note the correspondence of the 8-8 cluster solution, shown in blue, with the 2-2 cluster solution in Fig.~\ref{fig:3}(a).\\
There are two $\beta$ values, $\beta_1$ and $\beta_2$, between which the synchronized solution is unstable, see Fig.~\ref{fig:4}.
{These bifurcation points are transverse bifurcations~\cite{Banaji2002}, consisting of} transcritical bifurcations of all unbalanced cluster states and pitchfork bifurcations of the balanced state {(the latter only for an even number of total oscillators)}.
{We abbreviate this transverse bifurcation at the synchronized solution with $\mathbf{T}_s$.}
The two solution branches of each transcritical and of the pitchfork bifurcations {of the balanced solution}, respectively, connect the two bifurcation points at $\beta_1$ and $\beta_2$.
Each of the unbalanced clusters is born respectively destroyed in a saddle-node bifurcation {$\mathbf{SN}_{N-x, x}$},
the most outer one{,} corresponding to the \mbox{$(N-1)$-1} cluster state{,} being the only one that {produces} a stable branch.
The other cluster states, as well as the balanced one, are stabilized through further {transverse} bifurcations {$\mathbf{T}_{a|N-x, x}$} ({the resulting branches} not shown), compare Fig.~\ref{fig:3}(a),
{and destabilized by the transverse bifurcations $\mathbf{T}_{b | N-x, x}$ (the resulting branches also not shown)}.
In this way, two staircases of overlapping stable cluster states are generated, whereby the cluster distribution of the cluster states in subsequent steps differ by just one oscillator.
This leads to two cascades of transitions between the two synchronized regions.\\
If we start from the stable synchronized solution for $\beta<\beta_1$ and slowly increase $\beta$, the system goes from the synchronized solution to a 15-1 cluster state,
and then {to} a 14-2 state and so forth,
traversing a cascade up to the balanced 8-8 cluster and back,
until it settles again on the synchronized solution, see the blue curve in Fig.~\ref{fig:hyst_trans}.
Thereby, the originally larger cluster with amplitude $R_{C_1}$ (cf. Fig.~\ref{fig:4}(a)) becomes the smaller one with amplitude $R_{C_2}$ (cf. Fig.~\ref{fig:4}(b)) beyond the balanced cluster state.
The second staircase, and with it the hysteretic behavior, can be seen when we subsequently reduce $\beta$ again, cf. the orange curve in Fig.~\ref{fig:hyst_trans}.\\
Furthermore, from Fig.~\ref{fig:4}(a) it becomes obvious that all unbalanced cluster solutions, that is all cluster solutions except the 8-8 cluster,
bifurcate with the synchronized solution in a transcritical bifurcation.
In particular, the unbalanced cluster states exist on both sides of both bifurcation points where the synchronized solution changes stability.
{In addition}, all cluster solutions lose stability through the {transverse} bifurcations {$\mathbf{T}_{a|N-x, x}$} and {$\mathbf{T}_{b|N-x, x}$}(the resulting branches are not shown in Fig.~\ref{fig:4}(a)),
in which either the cluster with the {largest} or {smallest} amplitude breaks up.
This is true except for the most unbalanced, the 15-1, cluster,
for which the smallest cluster cannot break up.
There we find that this stable state is destroyed in {the} saddle-node bifurcation {$\mathbf{SN}_{N-1, 1}$} instead.
In addition, each kind of unbalanced cluster solution is stable in two different regions in parameter space. 
Each of these stable regions lies close but slightly shifted to the stable regions of neighboring cluster states.
\section{Persistence and Cluster Singularities}
The observed phenomena of slightly shifted bifurcations can be explained with the concept of persistence.
Loosely speaking, if two attractors are hyperbolic and close in phase space,
then bifurcations of those attractors are also close in parameter space~\cite{Banaji2002}.
In addition, one can infer that between the cluster states shown in Fig.~\ref{fig:4},
which must also exist for larger ensembles,
{these larger networks exhibit many more cluster states,}
and using persistence, their stability must be similar to that of the ones seen in Fig.~\ref{fig:4}.\\
This explains the cascade-like transition from balanced cluster states to the homogeneous solution in large ensembles,
where, when changing a parameter, one oscillator after another joins the other cluster until the synchronized solution is reached.
We conjecture that for infinitely large ensembles, the cluster attractors are infinitesimal close, and thus this process becomes continuous.\\
Turning back to Fig.~\ref{fig:2}(b), we observe that there is a codimension-2 point (pink point in Fig.~\ref{fig:2}(b)) where the stable 2-2 cluster {state} bifurcates {supercritically} into the synchronized solution in a pitchfork bifurcation.
\comment{This is in contrast to the phenomena observed in the literature, where the transition to the synchronized solution occurs via the unbalanced cluster solutions.}
For the Stuart-Landau ensemble, Eq.~\eqref{eq:sle}, this point can be found analytically as
\begin{align}
  \alpha_\text{CS} & = -\frac{1\pm \sqrt{3} c_2}{2},\\
  \beta_\text{CS} & = \frac{-c_2\pm \sqrt{3}}{2},
\end{align}
with the derivation shown in Appendix B.\\
The characteristics of such a point is that, when starting from a balanced 2-cluster solution and changing the parameters {across} this bifurcation point,
the two clusters approach each other and finally merge and form the synchronized solution.
This is what we call a \textbf{cluster singularity}.\\
\begin{figure*}[ht]
  \centering
  \includegraphics{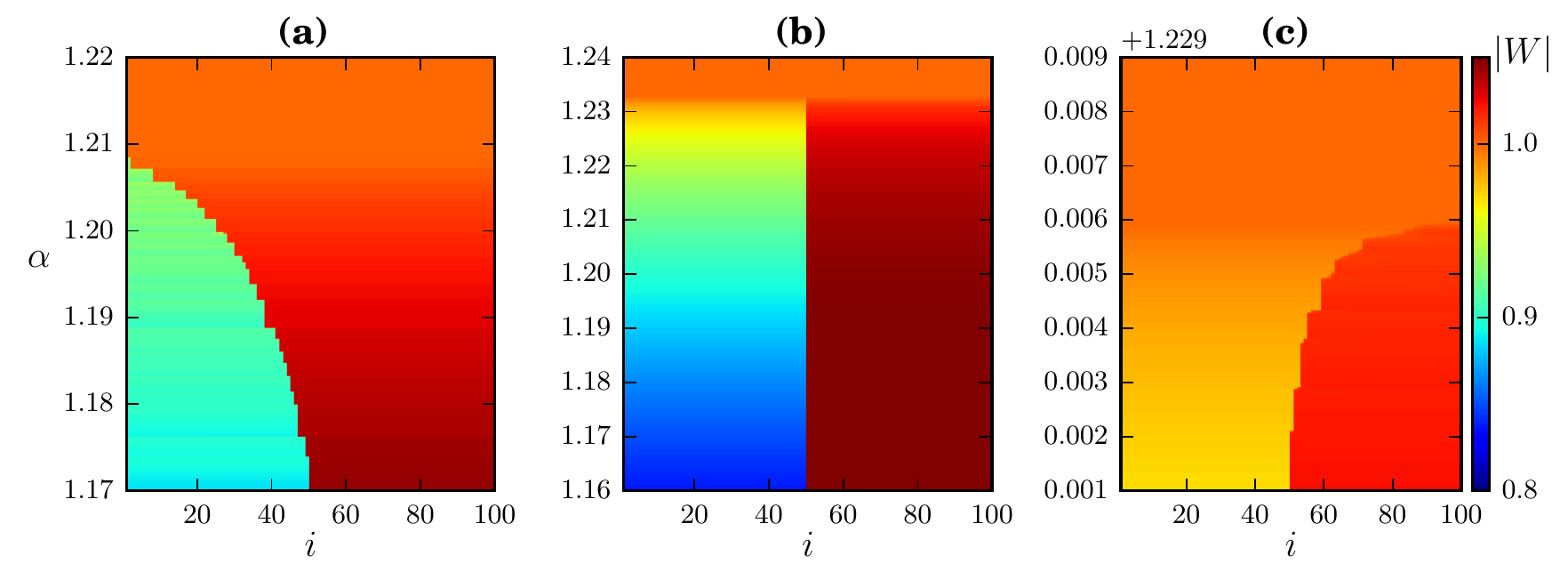}
  \caption{Simulations of the globally coupled Stuart-Landau ensemble close to the cluster singularity for $N=100$ oscillators and $c_2=2$,
    (a) $\beta=\beta_\text{CS}+0.3>\beta_\text{CS}$, (b) $\beta=\beta_\text{CS}$ and (c) $\beta=\beta_\text{CS}-0.05<\beta_\text{CS}$.
    The direction in which $\alpha$ is changed is from small to large values.}
\label{fig:5}
\end{figure*}
However, when varying the parameters such that one turns around the codimension-2 point either clock- or anticlockwise
(that is, changing $\alpha$ and $\beta$ along a path which circumvents the singularity
on the left or on the right, cf. pink arrows in Fig.~\ref{fig:2}(b)), then
either of the clusters shrinks and single oscillators join the other cluster until
all oscillators finally form the synchronized solution.
This scenario can be verified using numerical simulations and is shown in Fig.~\ref{fig:5}.
There, simulations of $N=100$ Stuart-Landau oscillators are shown when avoiding the cluster singularity anticlockwise (Fig.~\ref{fig:5}(a)),
when directly crossing over the cluster singularity (Fig.~\ref{fig:5}(b)) and when avoiding the cluster singularity in a clockwise manner (Fig.~\ref{fig:5}(c)).\\
The cluster singularity serves as an organizing center for nearby unbalanced cluster solutions.
Recall that all unbalanced 2-cluster solutions get destroyed in {the} saddle-node bifurcations {$\mathbf{SN}_{N-x, x}$}, cf. Fig.~\ref{fig:4}.
In the cluster singularity, all these saddle-node bifurcations as well as the {transverse} bifurcations {$\mathbf{T}_{a | N-x, x}$ and $\mathbf{T}_{b | N-x, x}$} that alter the
stability properties of the cluster states collapse to a single point in phase space, suggesting the name cluster singularity.
Note that when crossing the singularity, the stable balanced cluster solution directly bifurcates into the synchronized solution.
{
In Fig.~\ref{fig:9} we summarize the unfolding of the cluster singularity as suggested from the discussed bifurcation diagrams and simulations.
Fig.~\ref{fig:9}(a) shows the three ``primary'' bifurcations:
The middle, black curve is the transverse bifurcation $\mathbf{T}_{s}$, in which the synchronized solution becomes unstable and the balanced cluster state is born,
and all unbalanced cluster states merge in a transcritical bifurcation (cf. Fig.~\ref{fig:4}).
Note that all the cluster states participating in this transverse bifurcation are unstable.
The blue curve symbolizes the first of a cascade of saddle-node bifurcations $\mathbf{SN}_{N-x, x}$ creating pairs of unbalanced $(N-x)$-$x$ cluster solutions.
The first saddle-node bifurcation, $\mathbf{SN}_{N-1, 1}$, corresponds to the most unbalanced $(N-1)$-$1$ cluster state and involves a saddle and a stable cluster solution.
All the other saddle nodes generate two unstable cluster branches. 
The red curve is the last of a cascade of transverse bifurcations $\mathbf{T}_{a| N-x, x}$ through which the cluster solutions are stabilized, stabilizing the balanced $N/2$-$N/2$ cluster state
(cf. the 2-2 cluster state in Fig.~\ref{fig:3}(a), which gets stabilized through supercritical pitchfork bifurcations involving 2-1-1 cluster solutions).
The parameter regions in which the mentioned different bifurcation cascades occur are indicated in Figs.~\ref{fig:9}(b)~to~\ref{fig:9}(d).
The saddle node cascades $\mathbf{SN}_{N-x, x}$ are bounded by the first saddle-node bifurcation curve $\mathbf{SN}_{N-1, 1}$ and the transverse bifurcation $\mathbf{T}_s$ and are shown as hatched blue regions in Fig.~\ref{fig:9}(b).
The different slopes of the stripes on the two sides of the cluster singularity indicate that different cluster solutions are involved
(on one side the larger cluster has the larger amplitude, on the other side the larger cluster is the one with the smaller amplitude).
The cascades of transverse bifurcations $\mathbf{T}_{a|N-x, x}$
that stabilize the 2-cluster states by splitting off unstable 3-cluster branches take place in the green shaded areas in Fig.~\ref{fig:9}(c).
These cascades extend from close to the first saddle node curve $\mathbf{SN}_{N-1, 1}$ up to the transverse bifurcation curve $\mathbf{T}_{a|N/2, N/2}$ in which the balanced state is stabilized (red curve).
Again to the right and the left of the cluster singularities different cluster states are involved.
Finally, each branch of these stable 2-cluster states is destabilized by a second cascade of transverse bifurcations $\mathbf{T}_{b|N-x, x}$, again involving 3-cluster states.
The cascade starts between the red and the black curve on one side of the cluster singularity and extends to the red curve on the other side,
such that within the stability region of the balanced 2-cluster state the two cascades involving the different solution branches coexist (cf. Fig.~\ref{fig:9}(d)).}
\begin{figure*}[ht]
  \centering
  \includegraphics[width=\textwidth]{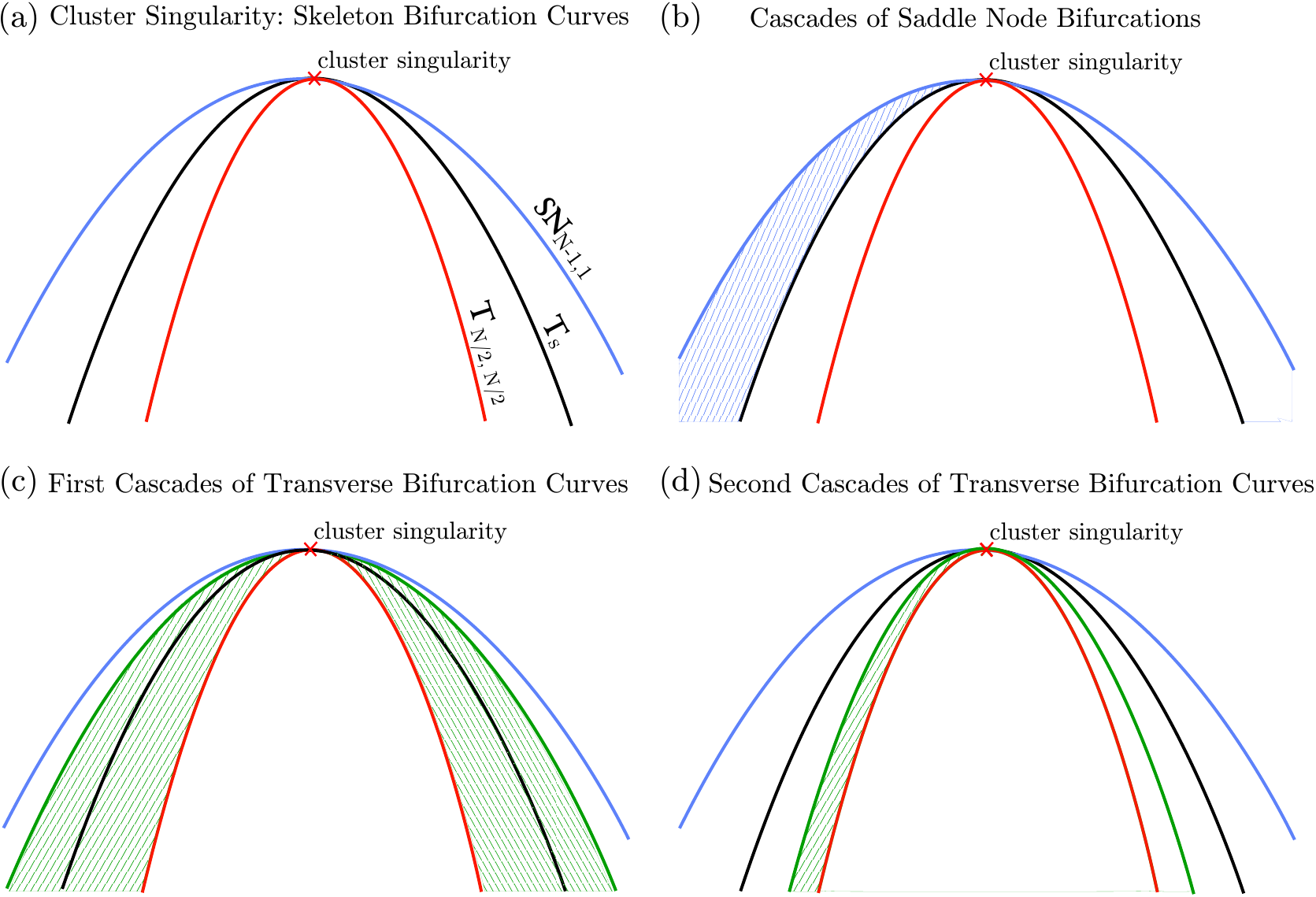}
  \caption{{Sketch of the bifurcation scenario close to the cluster singularity. The blue curve thereby represents the saddle node creating the most unbalanced cluster states.
    The black curve symbolizes the transverse bifurcation at the synchronized solution, $\mathbf{T}_s$, creating the unstable balanced cluster solution.
    The red curve represents the transverse bifurcation stabilizing the balanced cluster state, $\mathbf{T}_{N/2,N/2}$, cf. Fig.~\ref{fig:2}(b).
    For a detailed discussion of the different regions indicated in Figs.~(b)~to~(d) see text.}}
\label{fig:9}
\end{figure*}
{The regions filled with saddle nodes (cf. Fig.~\ref{fig:9}(b)), separating the region in which only the synchronized solution exists and where the balanced cluster solution exists, reminds of the Eckhaus instability~\cite{Eckhaus1965}.
There, the region between two parabolas marks the transition region in which more and more wave numbers become unstable~\cite{Tuckerman1990}.}\\

\section{Clustering in Spatially Extended Systems}
Adding a diffusive coupling in addition to the global coupling,
one obtains a globally coupled version of the complex Ginzburg-Landau equation~\cite{Battogtokh1996},
\begin{align*}
  \partial_t W & = W-\left(1+ic_2\right)\left|W\right|^2 W + d\vec{\nabla}^{\,2}W \\
  & + \left(\alpha+i\beta\right)\left(\int W\, dx-W\right),
\end{align*}
with $W=W(x,t)$ and $d\in\mathbb{C}$.
In a sense, such a system can be viewed as an ensemble of infinitely many oscillators, coupled locally and globally.
If the local coupling is weak ($\left|d\right|\ll 1$), then we expect the solutions of the Stuart-Landau ensemble to exist also in the spatially extended system.
For infinitely many oscillators, however, the 2-cluster solutions become infinitesimally close in phase space (cf. Fig.~\ref{fig:4} for 16 oscillators) and thus
infinitesimally small perturbations are sufficient to drive the solution from one cluster state to another.
This is also what we observe in numerical simulations: the diffusive coupling leads to the selection of a particular cluster distribution,
and the multistability of different cluster solutions,
as apparent in Fig.~\ref{fig:4} for Stuart-Landau oscillators, seems no longer to exist.
What is special about the then globally stable 2-cluster solution, however, still remains unknown.\\
\section{Conclusion}
To summarize our results, we have shown ways how clustering can occur in globally coupled ensembles of Stuart-Landau oscillators.
In particular, starting from small ensembles, we described how 2-cluster branches bifurcate, and extended this analysis to larger ensembles of oscillators.
Doing so, we found a codimension-2 point which we dubbed a cluster singularity: at this point, the stable balanced cluster solution bifurcates directly into the synchronized solution.
In addition, all saddle-node bifurcations generating unbalanced cluster solutions collapse in this point.
Using numerical simulations, we showed how ensembles of Stuart-Landau oscillators behave close to this cluster singularity{, thereby paving the way for a rigorous mathematical treatment}.
Since any oscillatory system close to the onset of oscillations can be mapped onto the dynamics of the Stuart-Landau oscillator,
we believe that cluster singularities are common in oscillatory systems with global coupling, and that an experimental observation of these should be possible.
Concludingly, we {glanced at} how our results extend to spatially extended systems, where the diffusive coupling seems to destroy the multistability.
{Yet, the relation of mean-field coupled ensembles of individual oscillators to spatially continuous oscillatory media with global, or long-range,
and diffusive coupling is still far from understood and poses an exciting challenge for future studies.
The same is true for secondary bifurcations of the 2-cluster states which are only point-wise known.
E.g.~in the regarded parameter window they may become unstable through supercritical Hopf bifurcations for smaller $\alpha$ values and even bifurcate into chimera states~\cite{Nakagawa1993, Kemeth2018}.}
\comment{We believe that our considerations may serve as a further step towards a better understanding of clustering behavior in coupled oscillators.
In addition, 2-cluster solutions in the regarded parameter windows may also become unstable through supercritical Hopf bifurcations for smaller $\alpha$ values, and even bifurcate into
chimera states.} How this transition occurs for different cluster distributions {constitutes another exciting and fundamental} question.

\acknowledgments{The authors thank Matthias Wolfrum, Oliver Junge, and Munir Salman for fruitful discussions.
  Financial support from the Deutsche Forschungsgemeinschaft (Grant no. KR 1189/18-1), the Institute of Advanced Study - Technische Universit\"{a}t M\"{u}nchen, funded by
  the German Excellence Initiative,
  and the Studienstiftung des deutschen Volkes is gratefully acknowledged.}

\appendix

\section{Balanced cluster solution}
In order to find the solutions of balanced 2-cluster states, it is sufficient to find the fixed points of the reduced 2-oscillator system
\begin{widetext}
\begin{align}
 \partial_t R_1 & = R_1-R_1^3 + \alpha\left(R_2\cos\left(\Delta \theta\right)-R_1\right) + \beta R_2\sin \left(\Delta \theta\right)\label{eqn:3_1}\\
 \partial_t R_2 &= R_2-R_2^3  + \alpha\left(R_1\cos\left(\Delta \theta\right)-R_2\right) - \beta R_1\sin \left(\Delta \theta\right)\label{eqn:3_2}\\
\partial_t\Delta\theta & = -c_2\left(R_1^2-R_2^2\right)+\beta\cos\left(\Delta\theta \right)\left(\frac{R_2}{R_1} -\frac{R_1}{R_2}\right)-\alpha\sin\left(\Delta\theta\right)\left(\frac{R_2}{R_1}+\frac{R_1}{R_2}\right)\label{eqn:3_3}.
\end{align}
\end{widetext}
In addition, these equations can be simplified by introducing the sum and the difference of the squared amplitudes, \(\gamma = R_1^2+R_2^2\) and \(\rho=R_1^2-R_2^2\), with
\begin{align*}
 \partial_t \gamma &= 2R_1\partial_t R_1 + 2R_2\partial_t R_2\\
 \partial_t \rho &= 2R_1\partial_t R_1 - 2R_2\partial_t R_2.
\end{align*}

This transforms Eqs.~\eqref{eqn:3_1} to~\eqref{eqn:3_2} into
\begin{widetext}
  \newpage
\begin{align*}
 \partial_t \gamma & = 2\left(1-\alpha\right)\left(R_1^2+R_2^2\right)-2{\left(R_1^2+R_2^2\right)}^2+4R_1^2R_2^2 + 4\alpha R_1R_2\cos\left(\Delta \theta\right) \\
 \partial_t \rho &= 2\left(1-\alpha\right)\left(R_1^2-R_2^2\right)-2\left(R_1^4-R_2^4\right)+4\beta R_1R_2\sin \left(\Delta \theta\right)\\
 \partial_t\Delta\theta & =-c_2\left(R_1^2-R_2^2\right)-\beta\cos\left(\Delta\theta \right)\frac{R_1^2 -R_2^2}{R_1R_2}-\alpha\sin\left(\Delta\theta\right)\frac{R_1^2+R_2^2}{R_1R_2}
\end{align*}
\end{widetext}
and using \(R_1R_2=\sqrt{\gamma^2-\rho^2}/2\),
\begin{align*}
 \partial_t \gamma & = 2\left(1-\alpha-\gamma\right)\gamma+\gamma^2-\rho^2 + 2\alpha\sqrt{\gamma^2-\rho^2}\cos\left(\Delta \theta\right)\\
 \partial_t \rho &= 2\left(1-\alpha-\gamma\right)\rho+2\beta \sqrt{\gamma^2-\rho^2}\sin \left(\Delta \theta\right)\\
  \partial_t\Delta\theta & =-c_2\rho-2\beta\cos\left(\Delta\theta\right)\frac{\rho}{\sqrt{\gamma^2-\rho^2}} \\
  &-  2\alpha\sin\left(\Delta\theta\right)\frac{\gamma}{\sqrt{\gamma^2-\rho^2}}.
\end{align*}
At a fixed-point solution, this system of equations must satisfy
\begin{align*}
  0 & = \left(1-\alpha-\gamma\right)\gamma+\frac{\gamma^2-\rho^2}{2} + \alpha\sqrt{\gamma^2-\rho^2}\cos\left(\Delta \theta\right)\\
  0 &= \left(1-\alpha-\gamma\right)\rho+\beta \sqrt{\gamma^2-\rho^2}\sin \left(\Delta \theta\right)\\
  0 & =-c_2\rho-2\beta\cos\left(\Delta\theta\right)\frac{\rho}{\sqrt{\gamma^2-\rho^2}} \\
  & - 2\alpha\sin\left(\Delta\theta\right)\frac{\gamma}{\sqrt{\gamma^2-\rho^2}}.
\end{align*}
Solving the first two equations for $\cos\left(\Delta\theta\right)$ and $\sin\left(\Delta\theta\right)$ yields
\begin{align}
  \sqrt{\gamma^2-\rho^2}\cos\left(\Delta \theta\right) & = -\left(1-\alpha-\gamma\right)\frac{\gamma}{\alpha}-\frac{\gamma^2-\rho^2}{2\alpha}\label{eq:rw1}\\
  \sqrt{\gamma^2-\rho^2}\sin \left(\Delta \theta\right)& = -\left(1-\alpha-\gamma\right)\frac{\rho}{\beta}\label{eq:rw2}
\end{align}
and inserted into the last equation,
\begin{widetext}
\begin{align*}
  0 &= -c_2\rho + \frac{\beta}{\alpha} \frac{\rho}{\gamma^2-\rho^2}\left[2\left(1-\alpha-\gamma\right)\gamma + \gamma^2-\rho^2\right] + 2\frac{\alpha}{\beta}\frac{\gamma}{\gamma^2-\rho^2}\left(1-\alpha-\gamma\right)\rho\\
\Rightarrow  0 &= -c_{2}\left(\gamma^2-\rho^2\right)+ \frac{\beta}{\alpha}\left[2\left(1-\alpha-\gamma\right)\gamma + \gamma^2-\rho^2\right] +2 \frac{\alpha}{\beta}\left(1-\alpha-\gamma\right)\gamma\\
\Rightarrow  0 &= -c_2\alpha\beta\left(\gamma^2-\rho^2\right) + 2\beta^2\left(1-\alpha-\gamma\right)\gamma + \beta^2\left(\gamma^2-\rho^2\right) + 2\alpha^{2}\left(1-\alpha-\gamma\right)\gamma\\
\Rightarrow  0 &= \left(\beta^2-c_2\alpha\beta\right)\left(\gamma^2-\rho^2\right) + 2\left(\alpha^2+\beta^{2}\right)\left(1-\alpha-\gamma\right)\gamma.
\end{align*}
\end{widetext}
Solving for $\rho^2$
\begin{equation}
  \rho^2= 2\frac{\alpha^2+\beta^{2}}{\beta^2-c_2\alpha\beta}\left(1-\alpha-\gamma\right)\gamma+\gamma^{2}.
\label{eqn:tsl_solcl}
\end{equation}
Using the identity $1= {\sin}^2\left(\Delta \theta\right)+{\cos}^2\left(\Delta \theta\right)$, we can write Eqs.~\eqref{eq:rw1} and~\eqref{eq:rw2}, yielding
\begin{align}
  \gamma^2-\rho^2 &={\left(1-\alpha-\gamma\right)}^2\frac{\rho^2}{\beta^2} \nonumber \\
  & + {\left(-\left(1-\alpha-\gamma\right)\frac{\gamma}{\alpha}-\frac{\gamma^2-\rho^2}{2\alpha} \right)}^2\nonumber\\
  \Rightarrow  \gamma^2-\rho^2 &={\left(1-\alpha-\gamma\right)}^2\frac{\rho^2}{\beta^2}\nonumber \\
  & + \frac{1}{4\alpha^2}{\left(\gamma^2-\rho^2+2\left(1-\alpha-\gamma\right)\gamma\right)}^2\label{eq:ref1}.
\end{align}
By inserting Eq.~\eqref{eqn:tsl_solcl} into Eq.~\eqref{eq:ref1}, we can solve it for $\gamma$ and obtain
\begin{widetext}
\begin{equation}
  \gamma = \frac{\left(1-\alpha\right)\left(3\beta-4\alpha c_2-\beta c_2^2\right)}{2\beta-4\alpha c_2-2\beta c_2^2} \pm \frac{\beta \sqrt{{\left(1-\alpha\right)}^2{\left(1+c_2^2\right)}^2 - 8 \beta^2\left(1-c_2^2\right) + 8\alpha c_2\left(3\beta -2\alpha c_2-\beta c_2^2\right)}}{2\beta-4\alpha c_2-2\beta c_2^2}.
  \label{eqn:cluster_2}
\end{equation}
\end{widetext}
Together with Eq.~\eqref{eqn:tsl_solcl}, this can be used to calculate $R_1$, $R_2$ and $\Delta \theta$.

\section{Cluster Singularities}

The idea is that at the cluster singularity, the saddle-node bifurcations of all unbalanced 2-cluster solutions hit the {transverse bifurcation} at which the
synchronous solution becomes unstable.
This must be true for any $\epsilon=N_1/N$.
Therefore, for simplicity, we take the limit $\epsilon \rightarrow 0$ in system Eqs.~\eqref{eqn:tsl_r1} to~\eqref{eqn:tsl_dp0}, yielding
\begin{align*}
  0 & = R_1-R_1^3\\
  0 & = R_2-R_2^3 + \alpha \left(R_1\cos\left(\Delta\phi\right) - R_2 \right) -\beta  R_1 \sin \left(\Delta\phi\right)\\
  0 & = -c_2\left(R_1^2-R_2^2\right)+\beta - \beta\cos\left(\Delta\phi \right)\left(\frac{R_1}{R_2}\right) \\
  & - \alpha\sin\left(\Delta\phi\right)\left(\frac{R_1}{R_2}\right).
\end{align*}
This means $R_1=1$ and thus leaves
\begin{align*}
  0 & = R_2-R_2^3 -\alpha R_2 + \alpha \cos\left(\Delta\phi\right)  -\beta \sin \left(\Delta\phi\right)\\
  0 & = -c_2\left(R_2-R_2^3\right)+\beta R_2- \beta\cos\left(\Delta\phi \right)-
  \alpha\sin\left(\Delta\phi\right).
\end{align*}
\begin{align*}
  R_2^2 & = -\frac{2 \alpha + 2 \beta c_2 -1-c_2^2}{2\left(1+c_2^2\right)}\\
  & \pm \frac{\sqrt{\left(2\alpha + 2\beta c_2 -1-c_2^2\right)^2 - 4\left(\alpha^2+\beta^2\right)\left(1+c_2^2\right)}}{2\left(1+c_2^2\right)}
\end{align*}
Setting $R_2^2=1$ means we are at the point where the cluster solution meets the synchronous solution ($\Delta \phi =0$ follows from $R_2=R_1=1$),
and from this the previous expression turns into
\begin{widetext}
\begin{align*}
  1 & = -\frac{2 \alpha + 2 \beta c_2 -1-c_2^2}{2\left(1+c_2^2\right)} \pm \frac{\sqrt{\left(2\alpha + 2\beta c_2 -1-c_2^2\right)^2 - 4\left(\alpha^2+\beta^2\right)\left(1+c_2^2\right)}}{2\left(1+c_2^2\right)}\\
\Rightarrow  2\left(1+c_2^2\right) & = -\left(2 \alpha + 2 \beta c_2 -1-c_2^2\right) \pm \sqrt{\left(2\alpha + 2\beta c_2 -1-c_2^2\right)^2 - 4\left(\alpha^2+\beta^2\right)\left(1+c_2^2\right)}
\end{align*}
\begin{align}
\Rightarrow  \left(2 \alpha + 2 \beta c_2 +1+c_2^2\right) & =  \pm \sqrt{\left(2\alpha + 2\beta c_2 -1-c_2^2\right)^2 - 4\left(\alpha^2+\beta^2\right)\left(1+c_2^2\right)} \nonumber\\
\Rightarrow  \left(2 \alpha + 2 \beta c_2 +1+c_2^2\right)^2 & =  \left(2\alpha + 2\beta c_2 -1-c_2^2\right)^2 - 4\left(\alpha^2+\beta^2\right)\left(1+c_2^2\right) \nonumber\\
\Rightarrow  \left(2 \alpha + 2 \beta c_2\right)^2 +\left(1+c_2^2\right)^2+2\left(2 \alpha + 2 \beta c_2\right)\left(1+c_2^2\right) & =  \left(2 \alpha + 2 \beta c_2\right)^2 \nonumber\\
& +\left(1+c_2^2\right)^2-2\left(2 \alpha + 2 \beta c_2\right)\left(1+c_2^2\right) \nonumber \\
  & - 4\left(\alpha^2+\beta^2\right)\left(1+c_2^2\right) \nonumber\\
\Rightarrow  4\left(2 \alpha + 2 \beta c_2\right)\left(1+c_2^2\right) & = - 4\left(\alpha^2+\beta^2\right)\left(1+c_2^2\right) \nonumber\\
\Rightarrow  \alpha^2+\beta^2 + 2\left(\alpha + \beta c_2\right) &= 0\label{eq:hom_instab}
\end{align}
\end{widetext}
which coincides with the curve at which the homogeneous solution becomes unstable.
For the saddle-node curve of the cluster solutions, the two solutions of $R_{2}^2$ must equal, and thus the discriminant must equal zero
\begin{widetext}
\begin{align*}
  0 & = \left(2\alpha + 2\beta c_2 -1-c_2^2\right)^2 - 4\left(\alpha^2+\beta^2\right)\left(1+c_2^2\right)\\
  4\left(\alpha^2+\beta^2\right)\left(1+c_2^2\right) & = \left(2\alpha + 2\beta c_2 -1-c_2^2\right)^2\\
  4\left(\alpha^2+\beta^2\right)\left(1+c_2^2\right) & = \left(2\alpha + 2\beta c_2\right)^2+\left(1+c_2^2\right)^2 -2\left(2\alpha + 2\beta c_2\right)\left(1+c_2^2\right)\\
  4\left(\alpha^2+\beta^2\right)\left(1+c_2^2\right) & = 4\left(\alpha + \beta c_2\right)^2+\left(1+c_2^2\right)^2 -4\left(\alpha + \beta c_2\right)\left(1+c_2^2\right).
\end{align*}
\end{widetext}
Now use that $-2\left(\alpha+\beta c_2\right) = \alpha^2 + \beta^2$ from Eq.~\eqref{eq:hom_instab} above,
\begin{align}
  4\left(\alpha^2+\beta^2\right)\left(1+c_2^2\right) & = 4\left(\alpha + \beta c_2\right)^2+\left(1+c_2^2\right)^2 \nonumber \\
  & -4\left(\alpha + \beta c_2\right)\left(1+c_2^2\right) \nonumber\\
  4\left(\alpha^2+\beta^2\right)\left(1+c_2^2\right) & = \left(\alpha^2+\beta^2\right)^2+\left(1+c_2^2\right)^2 \nonumber \\
  & +2\left(\alpha^2+\beta^2\right)\left(1+c_2^2\right) \nonumber\\
  0 & = \left(\alpha^2+\beta^2\right)^2+\left(1+c_2^2\right)^2 \nonumber \\
  & -2\left(\alpha^2+\beta^2\right)\left(1+c_2^2\right) \nonumber\\
  0 & = \left(\alpha^2+\beta^2 - 1-c_2^2\right)^2 \nonumber\\
  0 & = \alpha^2+\beta^2 - 1-c_2^2\label{eq:cluster_sn}
\end{align}
Eq.~\eqref{eq:cluster_sn} gives the saddle-node curve of the cluster with $\epsilon = 0$.
So for the cluster singularity, this saddle-node bifurcation coincides with the point at which the homogeneous solution becomes unstable, as given by Eq.~\eqref{eq:hom_instab},
which yields
\begin{align*}
  0 & = \alpha^2+\beta^2 + 2\left(\alpha + \beta c_2\right)\\
  0 & = \alpha^2+\beta^2 - 1-c_2^2.
\end{align*}
Subtraction of these two equations yields
\begin{align}
  0 & = 2\left(\alpha + \beta c_2\right)+1+c_2^2 \nonumber\\
  \alpha & = -\beta c_2 -\frac{1}{2}\left(1+c_2^2\right)\label{eq:alpha_cs}
\end{align}
and thus
\begin{align*}
  0 & = \left(\beta c_2 + \frac{1 + c_2^2}{2}\right)^2 + \beta^2 -  \left(1 + c_2^2\right)\\
  0 & = \left(2\beta c_2 + 1 + c_2^2\right)^2 + 4\beta^2 -  4\left(1 + c_2^2\right)\\
  0 & = 4\beta^2 c_2^2 + \left(1+c_2^2\right)^2 + 4\beta c_2 \left(1+c_2^2\right) + 4\beta^2\\
  & -  4\left(1 + c_2^2\right)\\
  0 & = 4\beta^2\left(1+ c_2^2\right) + \left(1+c_2^2\right)^2 + 4\beta c_2 \left(1+c_2^2\right)\\
  & -  4\left(1 + c_2^2\right)\\
  0 & = 4\beta^2 + \left(1+c_2^2\right) + 4\beta c_2   -  4\\
  0 & = \beta^2 + \beta c_2 + \frac{-3+c_2^2}{4}\\  
  \beta & = \frac{-c_2\pm \sqrt{3}}{2}.
\end{align*}
This solution plugged into Eq.~\eqref{eq:alpha_cs} yields
\begin{align*}
  \alpha & = -\frac{-c_2\pm \sqrt{3}}{2} c_2 -\frac{1}{2}\left(1+c_2^2\right)\\
   & = \frac{c_2\mp \sqrt{3}}{2} c_2 -\frac{1}{2}\left(1+c_2^2\right)\\
   & = -\frac{1\pm \sqrt{3} c_2}{2}.
\end{align*}
So, in total, we have at the cluster singularity
\begin{align}
  \alpha & = -\frac{1\pm \sqrt{3} c_2}{2}\\
  \beta & = \frac{-c_2\pm \sqrt{3}}{2}. 
\end{align}
This indicates two possible solutions for the cluster singularity.
Furthermore, it is worth mentioning that it seems to exist for all $c_2$ values.

\section{Numerical Methods}
For the integration of the Stuart-Landau ensemble, an implicit
Adams method with a fixed time step of $dt=0.01$ is used.
All figures are generated using matplotlib~\cite{Hunter2007}.

\bibliography{lit.bib}

\end{document}